\documentclass[prb,showpacs,twocolumn,aps,superscriptaddress,a4paper]{revtex4-1}
\usepackage{dcolumn,amssymb,amsmath,amsfonts,graphicx,latexsym,color}
\usepackage{epstopdf}
\begin{document}

\title{Phase diagram of a generalized off-diagonal Aubry-Andr\'{e} model with $p$-wave pairing}

\author{Tong Liu}
\affiliation{Department of Physics, Zhejiang Normal University, Jinhua 321004, China}
\author{Pei Wang}
\affiliation{Department of Physics, Zhejiang Normal University, Jinhua 321004, China}
\author{Shu Chen}
\affiliation{Beijing National Laboratory for Condensed
Matter Physics, Institute of Physics, Chinese Academy of Sciences, Beijing 100190,
China}
\affiliation{School of Physical Sciences, University of Chinese Academy of Sciences, Beijing, 100049, China}
\affiliation{Collaborative Innovation Center of Quantum Matter, Beijing, China}
\author{Gao Xianlong}
\affiliation{Department of Physics, Zhejiang Normal University, Jinhua 321004, China}

\date{\today}

\begin{abstract}
Off-diagonal Aubry-Andr\'{e} (AA) model has recently attracted a great deal of attention
as they provide condensed matter realization of topological phases. We numerically study
a generalized off-diagonal AA model with $p$-wave superfluid pairing in the presence of both commensurate and incommensurate hopping modulations. The phase diagram as functions of the
modulation strength of incommensurate hopping and the strength of the $p$-wave pairing
is obtained by using the multifractal analysis. We show that with the appearance of the $p$-wave pairing,
the system exhibits mobility-edge phases and critical phases with various
number of topologically protected zero-energy modes. Predicted topological
nature of these exotic phases can be realized in a cold atomic system of
incommensurate bichromatic optical lattice with induced
$p$-wave superfluid pairing by using a Raman laser in proximity to a molecular Bose-Einstein condensation.
\end{abstract}

\maketitle

\section{Introduction}
\label{n1}
Anderson localization, one of the famous quantum phenomena in condensed matter physics,
describes the absence of diffusion of matter waves due to interference in random
potentials~\cite{1an}. Although it has been proposed for nearly 60 years, direct observation of
this phenomena in a well controlled disorder has only been achieved in the ultracold atomic
experiments by Billy \emph{et al.}~\cite{8BILLY} and Roati \emph{et al.}~\cite{9ROATI}. Advances in the manipulation of ultracold
atoms offer a fascinating new laboratory for quantum simulating the solid-state models.
A notable example is the realization of one-dimensional (1D) Aubry-Andr\'{e} (AA) model of ultracold
atoms in an incommensurate quasiperiodic potential~\cite{9ROATI}.
Different from the conventional 1D Anderson model, where any amount of onsite-potential
disorder induces localization, the 1D AA model is an important paradigm which can host localized,
extended or critical eigenstates depending on the amplitude of the disorder strength~\cite{1PRB,2PRA,3PRA,He,Gramsch,7aubry}.

Different variations of the AA model have been studied~\cite{Biddle,xiaopeng,12PRL,13PRL}.
Based on a tight-binding model with finite next-nearest-neighbor hopping or by
including a long-range hopping term, the appearance of mobility edges is identified~\cite{Biddle}, which can
be precisely addressed by the duality symmetry~\cite{12PRL,13PRL}. The interplay of
disorder and superconductivity is studied in the AA model with $p$-wave pairing interaction~\cite{14PRL,15PRL,16PRB,Cao,peng},
which leads to a topological phase transition from the topological superconducting phase
to a topologically trivial localized phase when the strength of disorder increases over a critical value.

Meanwhile, further extensions of the AA model have been proposed to include both the commensurate and
incommensurate off-diagonal hopping modulations (dubbed the off-diagonal AA model)
which lead to Anderson localization and nontrivial zero-energy edge states~\cite{17PRL,18PRB,19PRB}. 
The interplay of commensurate and incommensurate modulations leads to the existence of nontrivial zero-energy edge states even the reflection symmetry is broken by the incommensurate modulations~\cite{19PRB}. Its phase diagram shows three distinct phases: the extended, the topologically-nontrivial, and topologically-trivial mixed phases (with a mixture of localized and critical eigenfunctions), respectively. The two extended phases are found related by a symmetry transformation~\cite{liu}.

The main motivation of this paper is to study the interplay between the off-diagonal disorder and the $p$-wave
superfluid pairing which leads to a rich phase diagram.
The generalized off-diagonal AA model with $p$-wave superfluid pairing is described by the following Hamiltonian:
\begin{equation}\label{eq:ham}
    \mathcal{ H}=\sum_{j=1}^{L-1} [ -t_j c_{j}^\dagger c_{j+1} +\Delta c_{j}^\dagger c_{j+1}^\dagger + H.c. ],
\end{equation}
where $L$ is the chain length and $c_j^\dagger$ ($c_j$) is the fermionic creation
(annihilation) operator at site $j$.  $t_j = t +\lambda_{j}+V_{j}$ is the
off-diagonal hopping amplitude between the nearest-neighbor sites,
where $t$, $\lambda_{j}=\lambda\cos(2\pi b{j})$
and $V_{j}=V\cos(2\pi\beta{j}+\phi)$ are the tunneling amplitude, the commensurate
and the incommensurate modulations, respectively. $\lambda$ and $V$ are the corresponding
modulation amplitudes, and the inverse wavelengths of commensurate and incommensurate
modulations are denoted by $b$ and $\beta$, respectively.
$\Delta$ is the $p$-wave superfluid pairing which is taken to be real.
Without loss of generality, we choose $b=1/2$, $\beta=(\sqrt{5}-1)/2$
and the phase in the incommensurate modulation $\phi = 0$ and set the on-site potential
to zero. In the following, $t = 1$ is set to the energy unit.

\begin{figure}
	\centering
	\includegraphics[width=0.5\textwidth]{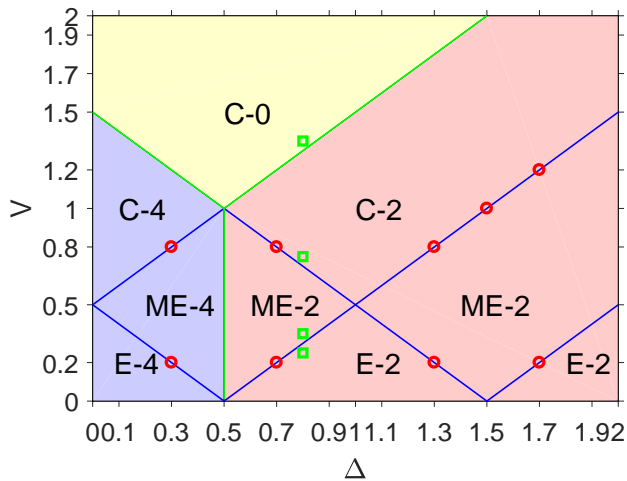}\\
	\caption{(Color online) Phase diagram of the off-diagonal AA model
	with the $p$-wave pairing strength $\Delta$ and the incommensurate modulation
	amplitude $V$. The commensurate modulation amplitude is set to $\lambda=0.5$.
	Nine different phases are separated by the blue solid and the green solid lines.
	The capital letter ``E'' denotes the extended phase,
	``ME'' denotes the mobility-edge phase, and ``C'' denotes the critical phase. The Arabic number
	after the capital letter is the number of zero-energy modes in this phase.
	All the phase boundaries, i.e. the blue solid and the green solid lines, are straight.
The colored region represents the phase of different topological zero-energy modes, separated by the green solid lines, where the energy gap closes at the boundaries.
	The red open circles are the boundary points whose positions
	are determined in Figs.~\ref{002} and~\ref{003},
	while the green squares refer to the values of $\left(\Delta,V\right)$ used in Fig.~\ref{004}.}
	\label{001}
\end{figure}

In the limit $\Delta=0$, this model reduces to the generalized off-diagonal
AA model introduced by Cestari \emph{et al.}.~\cite{19PRB}, where there exist
nontrivial zero-energy edge modes. Further study
points out that a large off-diagonal incommensurate modulation will drive
this system into a critical phase instead of a localized phase~\cite{liu}. In the
limit $\lambda=0$, the model describes a Kitaev chain with the off-diagonal incommensurate modulation
in the hopping amplitude but no on-site potentials, where a very
similar phase diagram as that for $\Delta=0$ exists. In this paper,
we consider both $\lambda\neq 0$ and $\Delta\neq 0$.

We find that the topological properties of the off-diagonal AA model
are profoundly affected by turning on the $p$-wave pairing.
We fix $\lambda=0.5$ throughout the paper. Our main results are summarized
in Fig.~\ref{001}. The extended phase, the mobility-edge phase
and the critical phase are separated by the blue solid lines.
For $0<V<0.5$, there are three regions which are in the extended phase.
And each of them from left to right hosts 4, 2 and 2 topologically protected
zero-energy modes, respectively. For an intermediate strength of $V$,
three regions in the mobility-edge phase host 4, 2 and 2 zero-energy modes, respectively.
For $V>0.5$, three regions in the critical phase are separated by the green solid lines
with the left (right) region hosting 4 (2) zero-energy modes.
The middle region is topologically trivial.

The rest of the paper is organized as follows. In Sec.~\ref{n2}, we
identify the transitions among the extended, the mobility-edge and the critical phases.
Sec.~\ref{n3} discusses the topological phase transition which is
signaled by the change of the number of zero-energy modes.
Finally, Sec.~\ref{n4} summarizes our results.

\section{Extended-Critical phase transition}
\label{n2}
Our numerical scheme of calculating the eigen wave function is as follows.
The Hamiltonian can be
diagonalized by using the Bogoliubov-de Gennes (BdG)~\cite{gnnes,lieb} transformation
\begin{equation}\label{}
\eta_n^\dagger = \sum_{j=1}^{L} [u_{n,j} c_{j}^\dagger + v_{n,j} c_{j}],
\end{equation}
where $u_{n,j}$ and $v_{n,j}$ denote the two components of
the wave-function at site $j$ which are real numbers.
The Schr\"odinger equation can be written as
\begin{equation}
\begin{split}
	& -t_{j-1}u_{n,j-1}- t_{j} u_{n,j+1} - \Delta (v_{n,j-1}- v_{n,j+1}) = E_n u_{n,j},\\
	& \Delta (u_{n,j-1}- u_{n,j+1}) + t_{j-1}v_{n,j-1}- t_{j} v_{n,j+1} = E_n v_{n,j},
\label{tb4}
\end{split}
\end{equation}
with $E_n$ denoting the eigenenergy. We obtain the eigen wave function
$\left(u_{n,1}, v_{n,1}, u_{n,2}, v_{n,2}, \cdots, u_{n,L}, v_{n,L}\right)^T$ by
diagonalizing the $2L \times 2L$ matrix
\begin{equation}\label{eq:BdGmatrix}
	\mathcal{H}=
	\begin{pmatrix}
	0 & A & 0 & \cdots & \cdots & \cdots & 0 \\
	A^\dagger & 0 & A & 0 & \cdots & \cdots & 0 \\
	0 & A^\dagger & 0 & A & 0 & \cdots & 0 \\
	\vdots & \ddots & \ddots & \ddots & \ddots & \ddots & \vdots \\
	0 & \cdots & 0 & A^\dagger & 0 & A & 0 \\
	0 & \cdots & \cdots & 0 & A^\dagger & 0 & A \\
	0 & \cdots & \cdots & \cdots & 0 & A^\dagger & 0,
	\end{pmatrix},
\end{equation}
where
\begin{equation}\label{}
A=\begin{pmatrix}
-(t+ \lambda_{j}+ V_{j}) & -\Delta \\
\Delta & t+ \lambda_{j}+ V_{j}
\end{pmatrix}.
\end{equation}

The inverse participation ratio (IPR) and the mean inverse participation ratio (MIPR) are usually
used to study the localization-delocalization transitions.
For a given normalized wave function ($\sum_{j=1}^{L}u_{n,j}^2 + v_{n,j}^2=1$), the IPR of the $n$-th eigenstate is defined as~\cite{27TH,28KO}
\begin{equation}
\text{IPR}_n =\sum_{j=1}^{L} (u_{n,j}^4 + v_{n,j}^4),
\end{equation}
which measures the inverse of the number of sites being occupied by atoms.
It is well known that the IPR of an extended state scales like $L^{-1}$ which goes
to $0$ in thermodynamic limit. But for a
localized state, since only finite number of sites are occupied,
the IPR is finite even in thermodynamic limit. For a critical state, the IPR scales as $L^{-\theta}$ with $0<\theta<1$.
The mean of IPR over all the $2L$ eigenstates is dubbed the MIPR which is expressed as
\begin{equation}
\text{MIPR}=\frac{1}{2L}\sum_{n=1}^{2L}\text{IPR}_{n}.
\end{equation}

\begin{figure}
	\centering
	\includegraphics[width=0.5\textwidth]{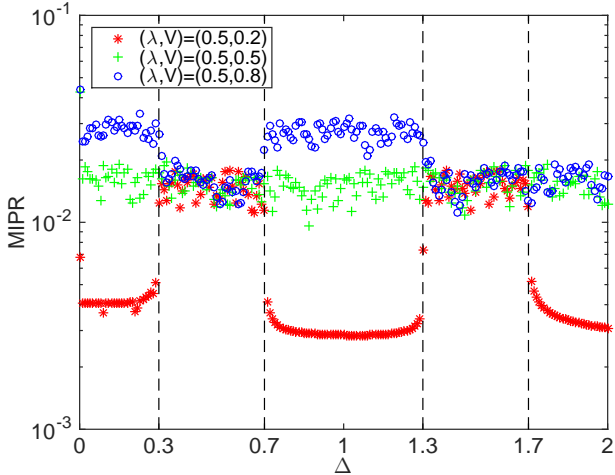}\\
	\caption{(Color online) MIPR as a function of $\Delta$ at different $V$ ($0<V<1$).
	The total number of sites is set to $L=500$.
	We use the open boundary condition in the calculation.
	The dashed lines show the value of $\Delta$ at which the MIPR changes abruptly.}
	\label{002}
\end{figure}

\begin{figure}
	\centering
	\includegraphics[width=0.5\textwidth]{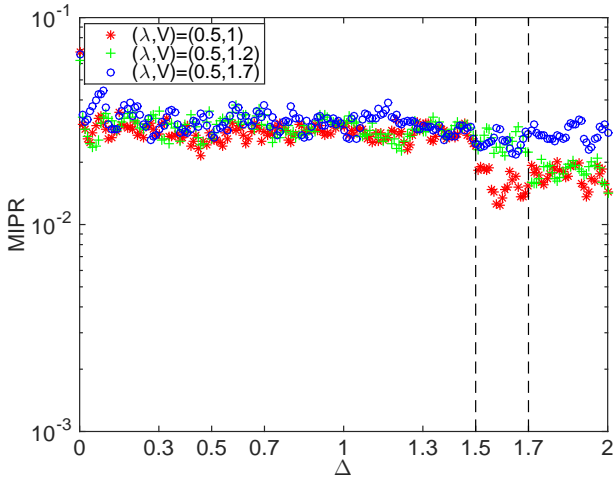}\\
	\caption{(Color online) MIPR as a function of $\Delta$ for different values
	of $V$ ($1<V<2$). The dashed lines show the value of $\Delta$ at which the MIPR changes abruptly.}
	\label{003}
\end{figure}
Figures~\ref{002} and~\ref{003} plot the MIPR as a function of $\Delta$ at different disorder amplitude $V$.
We set the chain length to $L =500$. We find that the MIPR changes abruptly at some values
of $\left(\Delta,V\right)$. These turning points signal a localization-delocalization transition,
which are marked by the red open circles in Fig.~\ref{001} located on the phase boundary.
We see that for $V=0.2$ (see Fig.~\ref{002}) the change of the MIPR at the turning points
is dramatic, because the transition there is from the extended phase to the mobility-edge phase.
For $V=0.8, 1$ and $1.2$ (see Figs.~\ref{002}-\ref{003}), the change of the MIPR at the turning points
is much smaller, as the transition is from the mobility-edge phase to the critical phase.
We have checked that, with increasing system's size $L$, the change of MIPR
at the turning points becomes even sharper. In the thermodynamic limit $L\rightarrow\infty$,
a discontinuity of the MIPR is expected, which signals a phase transition among
the extended, the critical, and the mobility-edge phases.
Notice that the MIPR distribution in Fig.~\ref{002} has three layers with the top
two layers staying close to each other. This corresponds to the three different phases.

\begin{figure}
	\centering
	\includegraphics[width=0.5\textwidth]{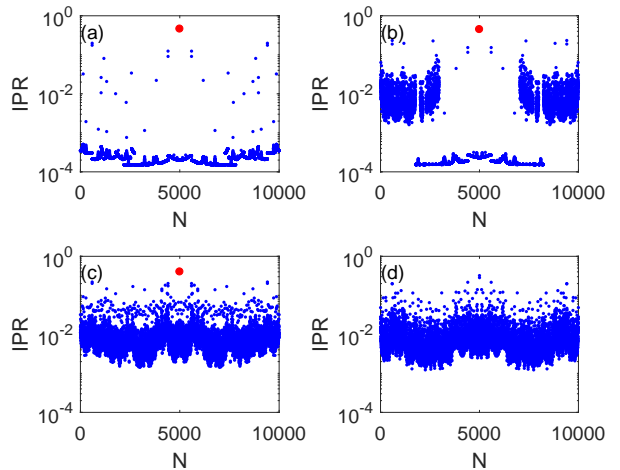}\\
	\caption{(Color online)  The distribution of IPRs over all the eigenstates at
	(a) $(\Delta,V)=(0.8,0.25)$ (the extended phase),
	(b) $(\Delta,V)=(0.8,0.35)$ (the mobility-edge phase), (c)
	$(\Delta,V)=(0.8,0.75)$ (the critical phase) and (d)
	$(\Delta,V)=(0.8,1.35)$ (the critical phase).
	The positions of $(\Delta,V)$ in different panels are
	marked by the green squares in the phase diagram Fig.~\ref{001}.
	The $x$-axis represents the eigenstate index with the corresponding
	eigenenergies increasing monotonically with increasing $N$.
	The number of sites is set to $L=5000$.
	The red dots represent the zero-energy modes, which are present in the
	topologically-nontrivial phase.}
	\label{004}
\end{figure}

The distribution of the IPR over different eigenstates is plotted
in Fig.~\ref{004} for $\Delta =0.8$. We find that there exist zero-energy modes (the red dots) in the region $V<1-\lambda+\Delta$
and no zero-energy modes otherwise. This confirms that the regions
below $V=1-\lambda+\Delta$ for $V>0.5$ are in the topologically-nontrivial phase,
while the region above $V=1-\lambda+\Delta$ is topologically trivial.
In Fig.~\ref{004}(a), the IPR of almost all the eigenstates are close to each other, being as low as $10^{-4}$.
This indicates that it has a pure energy spectrum with all the eigenstates being extended.

On the other hand, the critical phase has a significantly
different IPR distribution (see Fig.~\ref{004}(c) and~(d)). In these two panels,
the value of the IPR is at least one order of magnitude larger than that of
the extended state. At the same time, the IPRs of different eigenstates disperse widely
in a range over two orders of magnitude from $10^{-3}$ to
$10^{-1}$, indicating that they are also not the localized states.
The difference between Fig.~\ref{004}(c) and~(d) is that whether
they contain the zero-energy modes (the red dots) or not.

The most interesting phase emerging in our model
is the mobility-edge phase (see Fig.~\ref{004}(b)). As the eigenenergy increases,
at some specific energy the IPR suddenly jumps from a value around $10^{-2}$
(a typical value for the critical states as shown in the panels (c) and (d))
to $10^{-4}$ (a typical value for the extended states).
The jump of the IPR suggests that there exist the mobility edges in the energy spectrum.
The existence of the mobility edge also explains why the MIPR in Fig.~\ref{002} displays three layers.
In the mobility-edge phase, the extended eigenstates and the critical
eigenstates coexist leading to the MIPR which is between those
of the pure extended phase and the pure critical phase. The MIPR
among different phases satisfies
\[\text{MIPR}_{\text{E}}<\text{MIPR}_{\text{ME}}<\text{MIPR}_{\text{C}}.\]
We have checked that this relation is correct in the whole phase diagram.

To clarify the nature of different phases, we will analyze the eigen wave functions
by the multifractal analysis which has been applied in many disordered models, especially
the quasiperiodic models~\cite{1PRB,31HI,32WA}. We choose the total number of sites
to be $L=F_m$ where $F_{m}$ is the $m$-th Fibonacci number. The advantage of
this choice is that the golden ratio can be approximated by
$\beta = (\sqrt{5}-1)/2 =\lim_{m \rightarrow \infty} F_{m-1}/F_{m}$. The scaling index
$\gamma_{n,j}$ is inferred from the onsite probability $p_{n,j}=u_{n,j}^2 + v_{n,j}^2$ according to
\begin{equation}
p_{n,j} \sim \left(1/F_{m}\right)^{\gamma_{n,j}}.
\end{equation}
As the multifractal theory states, the number of sites whose scaling index
lies between $\gamma$ and $\gamma+d\gamma$ is proportional to $F_{m}^{f\left( \gamma\right)}$
in the scaling limit $m \to \infty$. For the extended wave functions, the largest onsite probability
scales as $\textbf{max}(p_{n,j}) \sim \left(1/F_{m}\right)^{1}$, indicating
$\gamma_{min}= 1$. On the other hand, for the localized wave functions, $p_{n,j}$ is
significant only at few sites but almost zero at the others, indicating
$\textbf{max}(p_{n,j}) \sim \left(1/F_{m}\right)^{0}$ and $\gamma_{min}= 0$.
For the critical wavefunctions, $f(\gamma)$ is a smooth function in the
interval $[\gamma_{min},\gamma_{max}]$ with $0<\gamma_{min}<1$.
Therefore, one can distinguish the extended, the localized and
the critical wave functions by calculating $\gamma_{min}$ which is defined as
$\textbf{max}(p_{n,j}) \sim \left(1/F_{m}\right)^{\gamma_{min}}$. Namely,
\begin{align}
&\hspace{0mm}\text{for the extended wave function,}~~~~
\gamma_{min}=1, \nonumber \\
&\hspace{0mm}\text{for the critical wave function,}~~~~
\gamma_{min}\ne 0,1, \nonumber \\
&\hspace{0mm}\text{for the localized wave function,}~~~~
\gamma_{min}=0.
\end{align}

\begin{figure}
	\centering
	\includegraphics[width=0.5\textwidth]{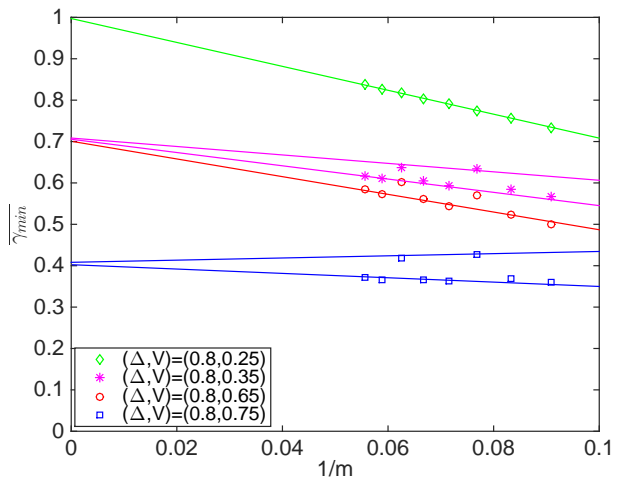}\\
	\caption{(Color online)  $\overline{\gamma_{min}}$ as a function of $1/m$ for
		$(\Delta,V)=(0.8,0.25)$, $(0.8,0.35)$, $(0.8,0.65)$ and $(0.8,0.75)$.
		These four points are located in the extended, the mobility-edge and the critical phases, respectively.}
	\label{005}
\end{figure}

For a lattice of $F_m$ sites, there exist $2F_m$ eigenstates.
We plot the average of $\gamma_{min}$ over all the eigenstates (denoted by $\overline{\gamma_{min}}$)
as a function of $ 1/m $ for different values of $(\Delta,V)$ in Fig.~\ref{005}.
We find that $\overline{\gamma_{min}}$ extrapolates to $1$ for
$(\Delta,V)=(0.8,0.25)$, indicating that it is in the extended phase.
But $\overline{\gamma_{min}}$ extrapolates to $0.4$ for $(\Delta,V)=(0.8,0.75)$ which
should then be in the critical phase.
Furthermore, $\overline{\gamma_{min}}$ extrapolates to $0.7$ for $(\Delta,V)=(0.8,0.35)$
and $(\Delta,V)=(0.8,0.65)$ in the mobility-edge phase, where the contributions to $\overline{\gamma_{min}}$
come from both the extended wave functions and the critical wave functions.

\section{Topological phase transition}
\label{n3}
In the above analysis, we identified the extended, the mobility-edge and the
critical phases and the boundaries between them in Fig.~\ref{001}.
Besides the localization-delocalization phase transitions, we also expect that
there exist topological phase transitions in our model.
In a very recent paper, Zeng et al.~\cite{zeng} studied a generalized
commensurate AA model with $p$-wave pairing and found different topological phases
including the Su-Schrieffer-Heeger-like (SSH-like)~\cite{su} topological phase
and the Kitaev-like~\cite{kitaev,wa} topological phase. In our model,
the off-diagonal incommensurate hopping $V$ leads to a more complicated phase diagram.
Since both SSH-like and Kitaev-like terms are included in the
Hamiltonian~(\ref{eq:ham}), we expect that, besides the SSH-like and the Kitaev-like phases,
new topological phases should emerge. We then check the energy spectrum of the system in
different regions of the phase diagram.

\begin{figure}
	\centering
	\includegraphics[width=0.5\textwidth]{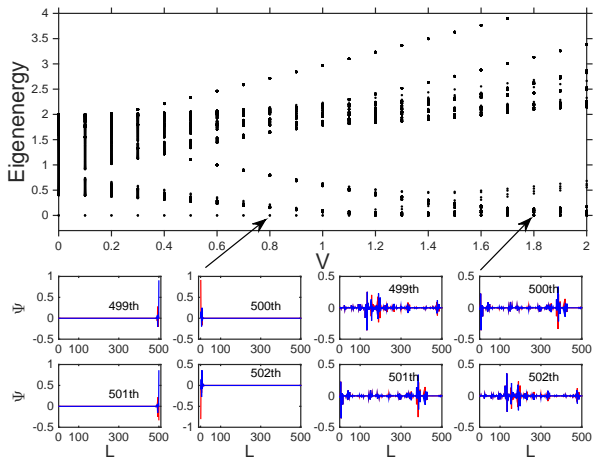}\\
	\caption{(Color online)  The top panel shows the spectra as a function of $V$ for $\Delta=0.3$.
The small eight panels in the bottom display the wave functions in the middle of the spectrum.
The blue and red solid lines represent the $u$ and $v$ components of the wave functions, respectively.
The bottom-left four panels are for $(\Delta, V)=(0.3, 0.8)$, while the bottom-right
four panels are for $(\Delta,V)=(0.3,1.8)$. The total number of sites is set to $L=500$.
}
	\label{006}
\end{figure}
\begin{figure}
	\centering
	\includegraphics[width=0.5\textwidth]{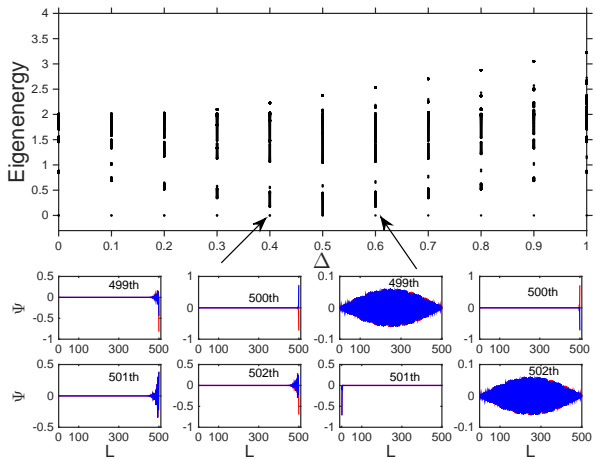}\\
	\caption{(Color online) The top panel shows the spectra changing with $\Delta$ at $V=0.3$.
The eight small panels are for the eigen wave functions in the middle of the spectrum.
The blue and red solid lines represent the $u$ and $v$ components, respectively.
The bottom-left four panels are for $(\Delta, V)=(0.4, 0.3)$, while the bottom-right
four panels are for $(\Delta,V)=(0.6,0.3)$.}
	\label{007}
\end{figure}

In Figs.~\ref{006} and~\ref{007}, we present the spectrum of the system
under the open boundary condition and find different number of zero-energy modes
for different $(\Delta, V)$. The change of the number of zero-energy modes
indicates a topological phase transition. In Fig.~\ref{006}, we see
the topologically nontrivial regime with zero-energy modes and the
trivial regime without. Furthermore, we check the eigen wave functions
at four different energies in the middle of the spectrum.
We clearly see that the wave functions of the zero-energy modes
are localized at the edge of the system (see the bottom-left panels of Fig.~\ref{006}).
There are four zero-energy edge modes at $(\Delta, V)=(0.3, 0.8)$,
indicating a topologically nontrivial phase. At $(\Delta,V)=(0.3,1.8)$
there exist no edge states, indicating a topologically trivial phase (see the bottom-right
panels of Fig.~\ref{006}). In Fig.~\ref{007},
the spectrum is gapped with four zero-energy modes as $\Delta<0.5$.
The gap closes at $\Delta=0.5$, and is reopened as $\Delta>0.5$ but
the number of zero-energy modes changes into two.

\begin{figure}
	\centering
	\includegraphics[width=0.5\textwidth]{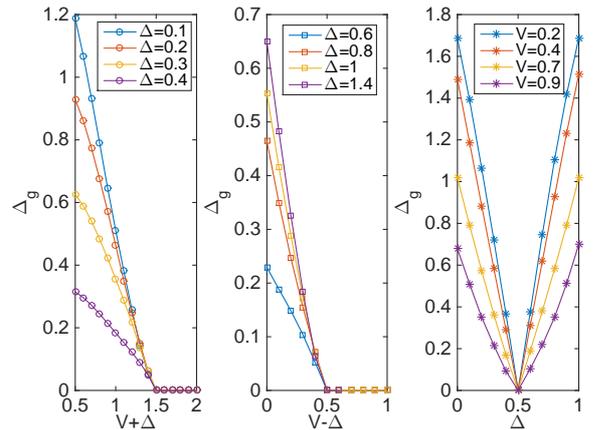}\\
	\caption{(Color online) $\Delta_g$ as a function of $V+\Delta$ (the left panel),
	$V-\Delta$ (the middle panel) and $\Delta$ (the right panel).
	The total number of sites is set to $L=500$. We choose the periodic boundary condition.
	}
	\label{008}
\end{figure}

The BdG matrix~(\ref{eq:BdGmatrix}) has the particle-hole symmetry,
therefore, the eigenenergies appear in pairs. We arrange the eigenenergies
in the ascending order and use $E_1$ to denote the smallest eigenenergy
which is larger than or equal to zero. $\Delta_g=2E_1$ can then be used to
explicitly determine the phase boundary between different topological phases.
In Fig.~\ref{008}, we plot $\Delta_g$ as a function of $V+\Delta$,
$V-\Delta$ and $\Delta$ for different values of $V$ or
$\Delta$ in the vicinity of the critical point.
Notice that the periodic boundary condition is adopted here, thereafter,
$\Delta_g$ indeed represents the energy gap since the edge modes are always absent.
The left panel shows that there is a finite gap as $V+\Delta$ is smaller than
a critical value $1.5$. When $V+\Delta$ exceeds $1.5$, the gap closes.
The middle panel shows that there is a finite gap when $V-\Delta$ is smaller than a critical
value $0.5$ but the gap also closes as $V-\Delta$ exceeds this critical value.
The right panel shows that the gap is closed at a critical value
$\Delta_c=0.5$ which is independent of $V$. As $\Delta$ exceeds $\Delta_c$, the gap is reopened.

We systematically check the gaps at different $V$ and $\Delta$ and find the
same behavior of the energy gap at $V = 1.5 - \Delta$, $V = 0.5 + \Delta$ and $\Delta = 0.5 $.
At these critical points, the gap reaches a minimum which approaches zero
in the limit $L \rightarrow \infty$. These three lines are in fact
the three green solid lines in Fig.~\ref{001}, which are the boundaries between different topological phases.

\section{Conclusions}
\label{n4}

In summary, we clarify the phase diagram of the off-diagonal
AA model with $p$-wave superfluid pairing in the presence of both commensurate and incommensurate hopping modulations. The localization and the topological properties
are greatly affected by the $p$-wave superfluid pairing. We have performed a systematic
numerical investigation of the phases of the systems. The IPR, MIPR, and the multifractal
analysis are used to identify the phase transitions among the extended, the mobility edge,
and the critical phases. To further identify the different topological properties,
the eigenenergies, the eigen wave functions and the energy gap are analyzed. We find that
the topological properties of the off-diagonal AA model are profoundly affected by the
induced nonzero $p$-wave pairing. A rich phase diagram is found due to the interplay
between the off-diagonal disorder and the $p$-wave superfluid pairing.
Two different types of extended and mobility-edge phase regions are identified,
each of them of 4 and 2 topologically-protected zero-energy modes, respectively.
For $V>0.5$, there are three critical phase regions in the phase diagram, two of them
of 4 and 2 zero-energy modes, respectively, and the third one is topologically trivial.

While the quasiperiodic disorder potentials can be realized in optical
lattices~\cite{8BILLY,9ROATI}, fast periodic modulations of interspecies
interactions allow the experimentalist to produce an effective model with small
diagonal and large off-diagonal disorder in cold atomic systems~\cite{Kosior}.
With ultracold atoms, it is also possible to induce directly superfluid $p$-wave pairing by using
a Raman laser in proximity to a molecular BEC~\cite{Jiang,Nascimbene,Hu2015}.
Combining these two techniques, it implies that experimental investigations of
the rich phase diagram due to the interplay between the off-diagonal disorder and the $p$-wave superfluid pairing
that we addressed here can be carried on.

\begin{acknowledgments}
This work was supported by
the NSF of China (Grant Nos. 11374266 and 11304280), and the Program for New Century
Excellent Talents in University, and the NSF of Zhejiang Province (Grant No. Z15A050001).

\end{acknowledgments}

\end{document}